\documentclass[5p,twocolumn]{elsarticle}

\usepackage{graphicx,epsfig}
\usepackage{amsmath,mathrsfs,amsfonts}
\usepackage {amssymb}
\usepackage {longtable}
\usepackage{multirow}
 \usepackage{enumerate}

\usepackage{bm}
\usepackage{amsfonts}
\usepackage{subfigure}
\usepackage{color}
\usepackage{epstopdf}
\usepackage{relsize}
 \usepackage[utf8]{inputenc}
\usepackage{hyperref}

\newcommand{\sch}{Schwarzschild}
\newcommand{\be}{\begin{equation}}
\newcommand{\en}{\end{equation}}
\newcommand{\bea}{\begin{eqnarray}}
\newcommand{\ena}{\end{eqnarray}}

\journal{Journal of \LaTeX\ Templates}

\biboptions{sort&compress}

\begin{document}

\begin{frontmatter}

\title{Poisson-Arago spot for gravitational waves}


\author{HongSheng Zhang}
\address{School of Physics and Technology, University of Jinan, Jinan, China}
\author{XiLong Fan\fnref{myfootnote}}
\address{School of Physics and Technology, Wuhan University, Wuhan 430072, China}
\fntext[myfootnote]{xilong.fan@whu.edu.cn}

%
%

\begin{abstract}
For the observer at infinity, a \sch~black hole serves as an  attractive opaque disk with a radius of 3$\sqrt{3} M$ that will produce the diffraction pattern of gravitational waves (GWs). In this study, we demonstrate that a bright spot, which is a diffraction effect analogous to the Poisson–Arago  spot in optics, will appear when an ingoing (quasi-)plane GW is diffracted by a \sch~black hole. Here, we propose the diffraction effect of the GWs described by the exact diffraction solution of the GWs using the Heun function. For the first time, the Fresnel half-wave zone method is proposed to calculate the angular part of the GW scattering stripes for the observer at infinity. The prospect of observing the diffraction bright spot is discussed with an eikonal approximation. For normal incidence (quasi)-plane waves with 100 Hz (0.1 Hz) frequency diffracted by the central black hole of the Milky Way, the time delay between the Earth bathed in a bright spot and the minimum of the first dark stripe is 3.86 (3860)  days. We will witness the second bright fringe (40$\%$ amplitude of the central bright spot) after 6.2 (6200) days. This new diffraction pattern involving the early phase of inspirals and pulsars as continuous gravitational wave sources is a potential scientific target for future space-and ground-based gravitational wave detectors, respectively.
\end{abstract}


\end{frontmatter}


\section{Introduction}\label{Introduction}
In 1916, Albert Einstein made the great prediction of gravitational waves (GWs) \citep{1916SPAW.......688E}. Since 2015,  these spacetime ripples have been detected by LIGO  and in conjunction with  Virgo \citep{2016PhRvL.116f1102A,2016PhRvL.116x1103A,2017PhRvL.118v1101A,2017ApJ...851L..35A,2017PhRvL.119n1101A,2017PhRvL.119p1101A}.  Diffraction and interference are the nature of  waves, as shown by pinhole diffraction and double-slit interference. The wave effects of GWs have been studied in the Newton–Coulomb potential limit in the context of strong lens \citep{1998PhRvL..80.1138N,1999PhRvD..59h3001B,2003ApJ...595.1039T}, where a gravitational  potential plays the role of an optical convex lens. Although no argument can be found on the wave nature of GWs confirmed by detections, the diffraction of GWs, being an important property of these waves, will be extremely interesting to observe experimentally.

In history, the Poisson–Arago spot appearing in the shadow of an opaque disk is an  ``experimentum crucis" to display the wave nature of light by showing the diffraction pattern. We will study the diffraction effects of GWs in a black hole background using the full theory of relativity, and show an effect similar to that of the Poisson–Arago spot.

The wave propagation in a weak gravitational potential is a general physical process studied using the optical approach \citep{1974PhRvD...9.2207P} with the Kirchhoff integral method (e.g., in terms of lens \citep{1998PhRvL..80.1138N,1999PhRvD..59h3001B}). Please see the review in \cite{2019RPPh...82l6901O}.  In the geometric limit, an Einstein ring, whose size is independent of the light wavelength, exists for a weak (Newtonian) gravitational potential  when the source lines up exactly in front of a bright background lens \cite{2019RPPh...82l6901O}.  The signal (photon or graviton) on the Einstein ring will be simultaneously detected by detectors. Given the poor sky localization ability of GW detectors  and the typical Einstein ring ($\theta_E \sim 1^{''}$), we argue that the Einstein ring of GWs, if it exists,  is unlikely to be observed in the near future.

The wave propagating in a strong field is investigated using the perturbation approach associated with the Regge–Wheeler equation \citep{1957PhRv..108.1063R} as a wave-scattering problem \citep{1968JMP.....9..163M,1970Natur.227..936V,1985ApJ...291L..33S,1977PhRvD..16.1636M,2008CQGra..25w5002D,2009PhRvL.102w1103C,2016CQGra..33g5011N,1992PhRvD..45.2609F,1994CQGra..11.2991A,1971ApJ...170L.105P,2019PhRvD.100f4037N}. Please see details in the review \cite{2000gr.qc....11025A} and the text book \cite{bhs}).
An intuitive geometric correspondence can be found between the high-frequency quasinormal modes (QNMs) of Schwarzschild black holes and null geodesics \citep{1972ApJ...172L..95G,2009PhRvD..79f4016C,2012PhRvD..86j4006Y}. Please see a review in \cite{Berti_note}.
The perturbation approach is usually divergent for an almost on-axis scattering angle $\sim 0 $ and $\pi$ and needs more approximate techniques (e.g., series reduction technique \cite{2008CQGra..25w5002D,1954PhRv...95..500Y,2009PhRvL.102w1103C} ) for some physical application and gravitational rainbow \citep{1985PhRvD..31.1869M,2019PhRvD.100b4007S}. As the exact solution of the Regge–Wheeler equation, the Heun function was recently used to study the QNMs outside the horizon \citep{2006CQGra..23.2447F}. It is interesting to use the Heun function to study the on-axis scattering in a strong field (e.g., diffraction bright spot of GWs). To the best of our knowledge, no literature has yet discussed the diffraction bright spot of GWs in terms of the Heun function.

In this study, we demonstrate that a \sch ~ black hole does similar work for GWs as an opaque disk in the Poisson–Arago effects in optics to estimate the measurement effect of the diffraction pattern of GWs, but with an attractive potential.
The environmental effects on the GW signal are insignificant within a broad class of scenarios \cite{2014PhRvD..89j4059B}. We ignore the lens environment and only use the \sch   ~ black hole potential in the wave equation (see Eq. (\ref{potential}). We also adopt an assumption that no interactions exist between waves with different polarizations, such that every graviton propagates independently. Under such an assumption, we use a scalar to replace a full tensor to describe the GW propagation. For the tensor nurture of the GWs in the optical approach, please see a recent work \cite{2019PhRvD.100f4028H} on the rotation of the polarization plane in a lensing system. We found that for sufficiently large distance $r$ and small angle approximation $\omega r\theta^2<<1$, the gravitational wave amplitude becomes
     \be \label{finaleq}
   \Phi \propto J_0\left(2\sqrt{\frac{M}{r}} \omega \Xi\right),
    \en
where $\omega$ is the gravitational wave frequency; $\theta$ is the forward scattering angle; $\Xi=r\theta$ is the radius coordinate in the polar chart at the image plane; $J_0$ is the Bessel function of the first kind (see more details in Eq.~(\ref{SA})).
Mathematically, the Poisson–Arago point of light in the flat space has a $\sim 1+J_1$ form. Here, we obtain a form proportional to $J_0$ for GWs. The gravitational potential plays a key role in this change, which attracts the gravity rays and leads to a more notable secondary maximum. Note that the bright spot, dark stripes, and bright fringes indicated in Eq. (\ref{finaleq}) should be treated as a pattern in a space screen and cannot be observed by a single gravitational wave detector at one time. Contrarily, the Einstein ring of lights is simultaneously observed by one optical telescope with sufficient angular resolution.

Although the observational angle is quite small, these spacetime stripes are a waveform -independent phenomenon and  could be a novel scientific target for future ground- and space-based gravitational wave detectors.

\section{Diffraction pattern of GWs by an opaque disk}

Without a back-reaction, the $l$-th partial wave of the gravitational wave $\hat{R}_{ls}$, which has an angular momentum $\hbar\sqrt{(l+1)l}$ relative to the scattering center, propagates on a \sch~ black hole background and is described by the wave equation as follows\citep{bhs}:
   \be
   \frac{\partial^2}{\partial t^2}\hat{R}_{ls}+(-\frac{\partial^2}{\partial x^2}+V_{ls})\hat{R}_{ls}=0,
   \en
 with effective potential $V_{ls}$,
    \be \label{potential}
    V_{ls}=(1-\frac{2M}{r})(\frac{2M(1-s^2)}{r^3}+\frac{l(l+1)}{r^2}+\mu^2),
    \en
where, $s$ is the spin particle for the graviton $s=2$. For the universality of this method, we introduce mass $\mu$ for the graviton. $t$ is the \sch~time coordinate. $x$ denotes the tortoise coordinate obtained by a coordinate transformation using the \sch~radius coordinate $r$ and the \sch~mass $M$,
 \be
  x=r+2M\ln(\frac{r}{2M}-1).
 \en

For a wave at a given frequency $\omega$, $\hat{R}_{ls}=R_{ls} e^{-i\omega t}$, the radius equation is presented as follows:
    \be
    (-\frac{\partial^2}{\partial x^2}-\omega^2+V_l)R_{ls}(r)=0.
    \label{me}
    \en
Eq. (\ref{me}) is a confluent Heun equation that has 24 independent local solutions, comprising a group that is isomorphic to the Coxeter group $D_3$.

The scattering boundary conditions in this case require
    \be
    \left|R_{ls}{(0)}\right|<\infty,~\left|R_{ls}{(2M)}\right|<\infty,~R_{ls}{(r\to\infty)}={R_{ls}}^\infty(r),
    \en
where, ${R_{ls}}^\infty(r)$ is a function of $r$ with a finite norm.
The boundary condition at $r=0$ is required to evade the Landau fall. The equation has a singularity at $r=2M$; thus, $R_{ls}{(2M)}$ must not diverge at this singularity. The third boundary condition at infinity is a general scattering state requirement.

The regular solution of the confluent Heun equation (Eq.~(\ref{me})) that satisfies the three boundary conditions simultaneously reads \citep{heun}
    \be
    R_{ls}(r)=K(r-2M)^ae^{\frac{-pr}{M}}{\rm Heunc}^{(a)}(p,\!~-b+a+1,\!~2a+1,~1,~d,~\frac{r}{2M}),
    \label{Rlsheun}
        \en
where,
   \bea
   a&=&2M(1-s^2)\omega i,\\
   b&=&-\left(2M\sqrt{\omega^2-\mu^2}+\frac{M\mu^2}{\sqrt{\omega^2-\mu^2}}\right)i,\\
   p&=&iM\sqrt{\omega^2-\mu^2},\\
   c&=&l(l+1)-8(\omega^2-\mu^2)M^2-6\mu^2M^2,\\
   d&=&c+2p(-2b+2a+1)-a(a+1).
      \ena
$K$ is a normalization constant. Heunc$^{(a)}$ is the angular generalized spheroidal function of the c-type, which is a special solution of the Heun equation (Eq.~(\ref{me})).

Note that,  this solution is a superposition of ingoing waves and outgoing waves, which in principle could be  used to  study  the interior of Schwarzschild black holes (see also discussion in \cite{2006CQGra..23.2447F}). Roughly speaking, it corresponds $sin(r)/r$ in the short-range interaction scattering problems.

The scattering stripes can be derived as follows \citep{bhs}:
   \be
    \Phi_l= \frac{1}{r} \sum_{l=2}^{\infty} (2l+1)e^{-i\omega t}P_l(\cos\theta)R_{ls}(r),
    \label{PHI}
    \en
where, $P_l(\cos \theta)$ is the normal Legendre polynomials. At the eikonal limit, the GWs can be treated as gravity rays (analogous to light rays). Rays with an impact parameter less than $3\sqrt{3}M$ fall into the \sch~black hole \citep{1992mtbh.book.....C}. Physically, they no longer make any contribution to the waves at infinity. At this point, a black hole serves as an opaque disk in the Poisson–Arago effects in optics. Note that, a black hole is almost a unique opaque disk that can shield GWs.

Owing to the oscillatory behavior of $\frac{R_{ls}}{r}$ at $r\to \infty$, the summation in Eq.~(\ref{PHI}) is generally divergent when $l$ goes infinity and cannot deal with the bright spot  effects of the GW unless arbitrarily with a cutoff at a larger $l$. We will introduce here a classical technique, namely, the Fresnel half-wave zone method, to avoid this difficulty in the next section. To the best of our knowledge, this approach is new to the gravitational scattering theory.

\section{Bright gravitational wave spot}
In principle, the gravitational wave amplitude at any spacetime point can be calculated using Eq.~(\ref{PHI}). However, the $R_{ls}$ formula in Eq.~(\ref{Rlsheun}) is complicated; thus, it is inconvenient to apply in astrophysics. We now discuss its asymptotic behavior at a large distance, which is useful for detection. First, we must confirm that it describes a (quasi)plane ingoing wave at $r\rightarrow \infty$. We create the following eikonal approximation,
    \be
    M\sqrt{\omega^2-\mu^2}>>1,
     \en
in which the angular quantum number $l$-th of a graviton can reach infinity satisfying $l>>1$. Therefore, the effects of the wave spin $s$ are negligible; that is, the scalar, vector, and tensor waves approximately obey the same equations.

\subsection{Radius part: Eikonal approximation}
At large distance and eikonal limit, $R_{ls}$ becomes \citep{heun}
     \begin{eqnarray}
     R_{ls}(r\to \infty)   \sim   \frac{1}{k}   \times \nonumber   \\
\exp\left(kr+2kM\ln\frac{r}{M}+\frac{M\mu^2}{k}\ln(\frac{r}{M}-1)+\delta_l-\frac{l\pi}{2}\right) ,
\end{eqnarray}
where, $\delta_l$ is the phase shift of the $l$-th partial wave, and we define $k^2=\omega^2-\mu^2$. We obtain the following phase shift at the eikonal limit
     \be
     \delta_l={\rm arg}\left[\Gamma(l+1-2ikM)\right]+2kM\ln 2kM,
     \en
by using the zero-order WKB approximation, where $\Gamma$ is the gamma function.

At the massless limit $\mu\to 0$, the scattering function $R_{ls}$ goes to
\begin{eqnarray}
      R_{ls}(r\to \infty,~\mu\to 0) \sim   \frac{1}{\omega } \times \nonumber\\
\exp\left(\omega r+2 \omega M\ln {2 \omega r}+{\rm arg}\left[\Gamma(l+1-2i \omega M)\right]-\frac{l\pi}{2}\right) .
\end{eqnarray}
We clearly return to the familiar result of the Newton–Coulomb potential scattering \citep{bhs}.

\subsection{Angular part: Fresnel half-wave zone method}\label{angular}
According to the Fresnel half-wave zone method, the first 1/4 zone presents full information of the diffracted waves, and the residues cancel each other out. We call such a 1/4 zone the effective region. This argument is also applicable to the gravity rays and any ray satisfying the superposition principle. For the angular momentum $J$ of a graviton with an angular quantum number $l$,
     \be
     J^2=l(l+1)\hbar^2.
     \en
By contrast, a graviton with an impact parameter $\rho$ has angular momentum,
     \be
     J=k\rho\hbar.
     \en
For the Fresnel-type diffraction, we have
     \be
     \rho_2^2-\rho_1^2=\lambda L,
     \en
where, $\lambda$ is the wavelength of the scattering wave; $\rho_1$  and $\rho_2$ are the radii of the interior and exterior edges of the half-wave zone, respectively;
and $L$ is the harmonic mean of the distances between the scattering center and the source and observation point.
The wave that can reach the observation point has a minimal $l_{\rm min}$,
     \be
     l_{\rm min}(l_{\rm min}+1)=k^2\rho_1^2.
     \en
For the \sch ~case, $\rho_1=3\sqrt{3}M$. $l_{\rm max}$ reaches the maximum value at the exterior edge of the effective zone.
     \be
      \sqrt{l_{\rm max}(l_{\rm max}+1)}=\frac{\pi L}{\rho_1+\sqrt{\rho_1^2+\lambda L}}.
      \en
Note that this method does not work for strong GWs, for which the back-reaction effect is so strong that the superposition principle fails.
We use the asymptotic formula of the Legendre function for a small angle diffraction:
     \be
     P_l(\cos \theta)=(\frac{\theta}{\sin \theta})^{1/2}J_0((l+\frac{1}{2})\theta).
     \en
\subsection{Small forward scattering amplitude}
In summary, one will detect the diffraction pattern amplitude of the GWs (Eq.~(\ref{PHI})) under the condition of a sufficiently large $r$ and a small forward scattering angle  approximation that becomes
     \be
     \Phi_l=e^{-i \omega t}e^{\omega r(1- \frac{\theta^2}{4})}e^{\pi M \omega}\Gamma(1-2i \omega M)J_0\left(2\sqrt{\frac{M}{r}} \omega \Xi\right),
     \label{SA}
     \en
where, $\Xi=r\theta$ is the radius coordinate in the polar chart at the image plane. Compared with Eq.~(\ref{PHI}), the above equation is greatly simplified. In the language of wave optics, considering paraxial rays as for the Poisson–Arago effect is sufficient.

This result seems similar to that of the scalar scattering image \citep{2016CQGra..33g5011N}. This is not surprising because only gravitons with a large $l$ have effects on the image. For such a graviton, the coupling result of the orbital and spin angular momenta is effectively equal to that of the orbital angular momentum. At the eikonal limit, massless particles with different spins obey the same equation of motion. A technical detail in our demonstration is that we do not need to replace $l(l+1)$ with $(l+\frac{1}{2})^2$ by hand in Eq.~(\ref{PHI}) to avoid divergence.

\section{Implication and discussion}
At the eikonal limit, the diffraction wave amplitude at a large angle is much lower than that of the forward scattering waves. Therefore, only the diffraction fringes of the GWs from a source located within a scatter angle (i.e., an angle between the propagate and scatterer (opaque disk) directions of the GWs) could be detected. The starting point of the angle estimation here is different from that of the single-slit diffraction of GWs if it can be realized. In this example, the small angle approximation (Eq.~(\ref{SA})) is applied. When one needs the full information of the diffraction waves in the total space, one has to return to the full form of the wave (Eq.~(\ref{PHI})). If an error of 1\% or 5\% of Eq.~(\ref{SA}) compared with the exact form (Eq.~(\ref{PHI})) is permitted, the cone’s apex angle is $7.1^0$ and $13.9^0$, corresponding to $1.5\%$ or $3\%$ of all-sky, which means that only a few parts of the gravitational wave source population could observe the diffraction bright spot of GWs.

We use the eikonal limit; thus, the upper bound of the wavelength in our calculation is approximately the radius of the ``opaque disk,"  which is $6\times 10^{10} m$ for the  supermassive black hole in our galaxy and corresponds to a frequency of 0.005~Hz. Therefore, planned space-based detectors, such as LISA \footnote{https://www.elisascience.org/}, Taiji \citep{2011CQGra..28i4012G}, and Tianqin \citep{2016CQGra..33c5010L}, could observe this type of event. The wavelength in our calculation has no theoretical lower limit. The realistic lower bound only depends on sensitivity gravitational wave detectors. It is $\sim $ a few hundred Hz for ground-based interferometer detectors.

The time delay between observing the bright spot and the dark stripe depends on the black hole mass ($M$), the relative position of the system ($r$ and $\theta$), and wavelength ($k$), as indicated in Eq.~(\ref{SA}). At first sight, one may think that the first term, which is the exponential term $e^{-\theta^2/4}$, will significantly affect the wave function strength. However, we will show that its impact is extremely weak and can be safely omitted in realistic cases. The principal property of the fringe is determined by the sector of the Bessel function $J_0$.

\begin{figure}
\centering
\includegraphics[width=7cm]{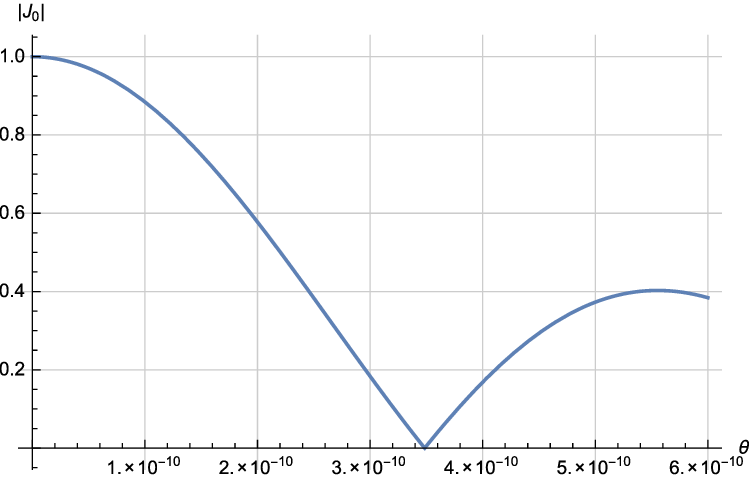}
\caption{The dark fringe appears at $\theta=3.485\times 10^{-10}$. The secondary bright fringe appears at $\theta=5.553\times 10^{-10}$.}
\label{X1}
\end{figure}

\begin{figure}
\centering
{\includegraphics[width=8cm]{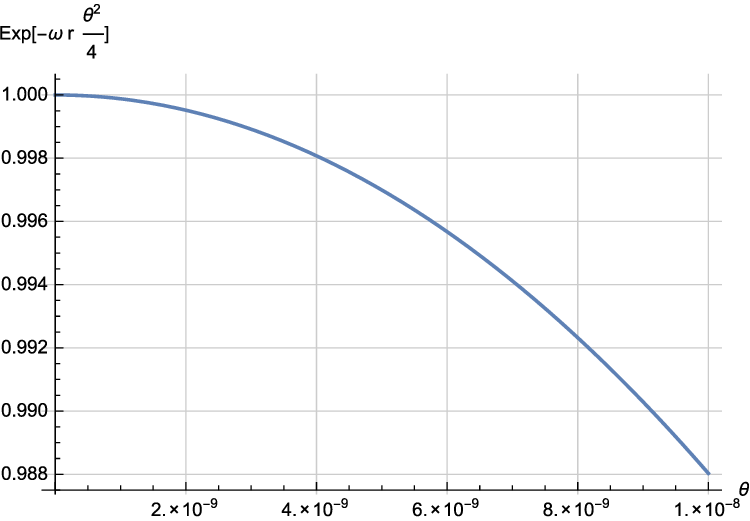}}
\caption{When $\theta<10^{-9}$, the exponential term is equal to 1 and has no effect on the fringe.}
\label{X2}
\end{figure}

We study here the most promising event, that is, a wave is diffracted by the central black hole of the Milky Way. Assume that the three points, namely, the source of the gravitational wave, the central black hole of the Galaxy, and the Earth, are collinear. We only live on the gravity diffraction axis and are now in a central bright spot of a gravitational wave. The numerical factor setting scales correspond to $M=4\times10^6 M_{\odot}$, $r=2.3\times 10^{20} m$, and $c=3\times 10^8 m/s$, and the velocity of the Sun in the Galaxy is 240 $ km/s$. For a 100 Hz-frequency wave, $k=2\pi\times 100/c$. Using the properties of the Bessel function (Fig. \ref{X1}), the first dark fringe appears at $\theta=3.485\times 10^{-10}$, whereas the next bright fringe appears at $\theta=5.553\times 10^{-10}$. Thus, the Earth floating from the central bright spot to the first dark fringe needs a trip that is $A_1=r\theta=2.3\times 10^{20}\times 3.485\times 10^{-10}=8\times 10^{10} m$. Substituting the velocity of the Earth in the Milky Way (i.e., 240 $ km/s$), we obtain the time needed to reach the first dark fringe as $8\times 10^{10}/(24\times 3600\times 240\times 10^3)$=3.86 days. We will reach the next bright fringe, whose luminosity is $40\%$ of the central bright spot, at $6.2$ days. Fig. \ref{X2} clearly showshows the effects of the exponential term to the fringes. For the 0.1 Hz GW, we will witness the minimum of the first dark stripe in 3860 days and experience the secondary maximum in 6200 days.

\section{Conclusion and discussion}
The proposed diffraction bright spot of the GWs is a waveform-independent phenomenon because we use the time-independent scattering theory. One only needs to search for the amplitude pattern according to Eq.~(\ref{SA}). To observe the bright spot, dark stripes, and bright fringes with planned gravitational wave detectors, we need a relatively stationary source, for example, a compact binary system in the early phase of inspiral for space-based detectors,  pulsars as continuous gravitational wave sources for ground-based detectors \cite{2021arXiv210700600T}, or binary-extreme-mass-ratio inspirals for both space- and ground-based detectors \citep{2018CmPhy...1...53C}. One must consider time-dependent scattering theories to study the waves from fast-changing sources (e.g., binary stars that are about to merge or are merging). (Supermassive) black holes are usually hosted in galaxies with deep potential. The lensing effect by a black hole host galaxy with the geometric optic limit (e.g., Einstein ring of GWs) can also be important for the detection data analysis. In future studies, we hope to investigate such a geometric optic limit effect by a black hole host galaxy and wave effect by a black hole.



\section*{Acknowledgements}
The authors would like to thank the referees for their
valuable comments which considerably improved the original
text. The authors thanks  Y. Chen for valuable comments.
This work was supported by  the National Natural Science Foundation of China under Grants  No. 11922303, 11673008, 11075106, 11575083 and 11275128.
X. F. was also supported by Hubei province Natural Science Fund for the Distinguished Young Scholars (No.~2019CFA052) and Newton International Fellowship Alumni Follow on Funding.


\bibliographystyle{spphys}

\bibliography{gw_sspc_v2_editor.bbl}   

\end{document}